\documentclass[preprint]{uai2026} 
                        
\usepackage[american]{babel}

\usepackage{natbib} 
\usepackage{mathtools} 
\usepackage{booktabs} 
\usepackage{tikz} 
\usetikzlibrary{positioning}
\usepackage{multirow} 
\usepackage{amssymb} 
\usepackage[ruled,vlined]{algorithm2e}
\usepackage{array}

\usepackage{makecell}
\usepackage{pgfmath}
\usepackage[table]{xcolor}
\usepackage[most]{tcolorbox}
\usepackage{caption}

\definecolor{prismLink}{HTML}{1C6E8C} 
\definecolor{prismCite}{HTML}{8E4B6A} 
\definecolor{prismURL}{HTML}{0F766E}  
\definecolor{prismFile}{HTML}{4C566A} 
\hypersetup{
  colorlinks=true,
  linkcolor=prismLink,
  citecolor=prismCite,
  urlcolor=prismURL,
  filecolor=prismFile,
  pdfborder={0 0 0},
  breaklinks=true
}

\definecolor{mygreen}{RGB}{47,150,98}
\definecolor{myred}{RGB}{200,80,80}
\definecolor{myblue}{RGB}{70,130,180}

\definecolor{phaseBorder}{RGB}{28,110,140}
\definecolor{phaseA}{RGB}{232,243,250}
\definecolor{phaseB}{RGB}{234,247,240}
\definecolor{phaseC}{RGB}{250,241,226}
\newsavebox{\phaseboxbox}
\newenvironment{phasebox}[1]{%
  \par\smallskip\noindent
  \def\phaseboxfill{#1}%
  \setlength{\fboxsep}{3pt}%
  \begin{lrbox}{\phaseboxbox}\begin{minipage}{0.99\linewidth}\vspace{1pt}%
}{%
  \end{minipage}\end{lrbox}%
  \begin{tikzpicture}
    \node[rounded corners=2pt, draw=phaseBorder, line width=0.6pt,
      fill=\phaseboxfill, inner xsep=4pt, inner ysep=3pt] {\usebox{\phaseboxbox}};
  \end{tikzpicture}%
  \par\smallskip
}

\newsavebox{\algoboxbox}
\newenvironment{algoblock}[1]{%
  \par\noindent
  \def\algoblockfill{#1}%
  \setlength{\fboxsep}{3pt}%
  \begin{lrbox}{\algoboxbox}\begin{minipage}{0.9\linewidth}\vspace{2pt}%
}{%
  \vspace{2pt}\end{minipage}\end{lrbox}%
  \begin{tikzpicture}
    \node[rounded corners=2pt, fill=\algoblockfill, inner xsep=4pt, inner ysep=3pt]%
      {\usebox{\algoboxbox}};
  \end{tikzpicture}%
  \par
}

\AtBeginDocument{\hbadness=10000 \vbadness=10000 \hfuzz=1pt \vfuzz=1pt}

\newcommand{\pmstack}[3]{\ensuremath{#1 {\scriptstyle #3}} \\ \ensuremath{\pm {\scriptstyle #2}}}

\def\cboxmax{10} 
\newcommand{\cbox}[4]{%
  \begingroup
  \pgfmathsetmacro{\cboxratio}{min(abs(#3)/\cboxmax,1)}%
  \pgfmathsetmacro{\cboxg}{1 - 0.30*\cboxratio}%
  \pgfmathsetmacro{\cboxb}{1 - 0.30*\cboxratio}%
  \ifdim #3 pt < 0pt
    \definecolor{cboxcolor}{rgb}{1,\cboxg,\cboxb}%
  \else
    \definecolor{cboxcolor}{rgb}{\cboxg,1,\cboxb}%
  \fi
  \tikz[baseline=(c.base)]%
    \node[rounded corners=1pt, fill=cboxcolor, inner xsep=2pt, inner ysep=1pt] (c)%
    {\makecell{\pmstack{#1}{#2}{#4}}};%
  \endgroup
}

\newcommand{\cboxours}[4]{%
  \begingroup
  \pgfmathsetmacro{\cboxratio}{min(abs(#3)/\cboxmax,1)}%
  \pgfmathsetmacro{\cboxg}{1 - 0.45*\cboxratio}%
  \pgfmathsetmacro{\cboxb}{1 - 0.45*\cboxratio}%
  \ifdim #3 pt < 0pt
    \definecolor{cboxcolor}{rgb}{1,\cboxg,\cboxb}%
  \else
    \definecolor{cboxcolor}{rgb}{\cboxg,1,\cboxb}%
  \fi
  \tikz[baseline=(c.base)]%
    \node[rounded corners=1pt, draw=black!20, line width=0.2pt, fill=cboxcolor, inner xsep=2pt, inner ysep=1pt] (c)%
    {\makecell{\pmstack{#1}{#2}{#4}}};%
  \endgroup
}

\title{Unifying Temporal and Structural Credit Assignment in LLM-Based Multi-Agent Prompt Optimization}

\author[1]{Wenwu~Li}
\author[1]{Yuran~Song}
\author[2]{Mingze~Zhao}
\author[1]{Bo~Jin}
\author[1]{Wenhao~Li\thanks{Corresponding author.}}

\affil[1]{%
    Tongji University\\
    Shanghai, China\\
    \texttt{\{wenwu,2250753,bjin,whli\}@tongji.edu.cn}
}
\affil[2]{%
    The University of Hong Kong\\
    Hong Kong, China\\
    \texttt{zhaomingze@connect.hku.hk}
}
  
\begin{document}
\maketitle

\begin{abstract}


While Multi-Agent Systems (MAS) empower Large Language Models to tackle complex reasoning tasks through collaborative interaction, optimizing their dynamics remains a formidable challenge due to the discrete, non-differentiable nature of the computation graph and the sparsity of global supervisory signals. 
Existing black-box optimizers struggle to attribute trajectory-level failure to specific local components, resulting in inefficient, high-variance exploration.
We argue that tractable MAS optimization needs \textbf{structural inductive biases} to disentangle error signals.
We propose \textbf{temporal and structural credit assignment}, which decomposes the objective along two axes: (i) \emph{temporal credit}, using state-space bottlenecks to identify critical rounds, and (ii) \emph{structural credit}, using stationary role policies to isolate agent contributions.
Leveraging these decomposed signals, we introduce a discrete, verbalized \textbf{block coordinate descent} algorithm for iterative refinement.
Rather than indiscriminate global updates, it alternates between optimizing role prompts and aggregation protocols, using LLM-generated ``proxy gradients'' to target only the identified weak links.
Across diverse reasoning benchmarks, our approach substantially reduces query complexity while improving performance, providing a principled and interpretable path toward self-improving MAS.

\end{abstract}

\section{Introduction}\label{sec:intro}

Large Language Models (LLMs) have evolved from static text generators into dynamic reasoning engines, achieving remarkable success in mathematical reasoning, code generation, and complex planning \citep{NEURIPS2020_1457c0d6, 10.5555/3600270.3602070}.
To transcend the limitations of single-model inference, recent research has shifted toward MAS, where specialized agents collaborate via iterative critique-and-revision cycles \citep{10.5555/3692070.3694667, zhang2025reinforcement}.
By distributing tasks across diverse roles (e.g., proposer, debater, synthesizer), MAS frameworks can theoretically solve problems that are intractable for a single monolithic model \citep{li-etal-2024-improving-multi}.

\begin{figure*}[htb!]
    \begin{center}
    \centerline{\includegraphics[width=\textwidth]{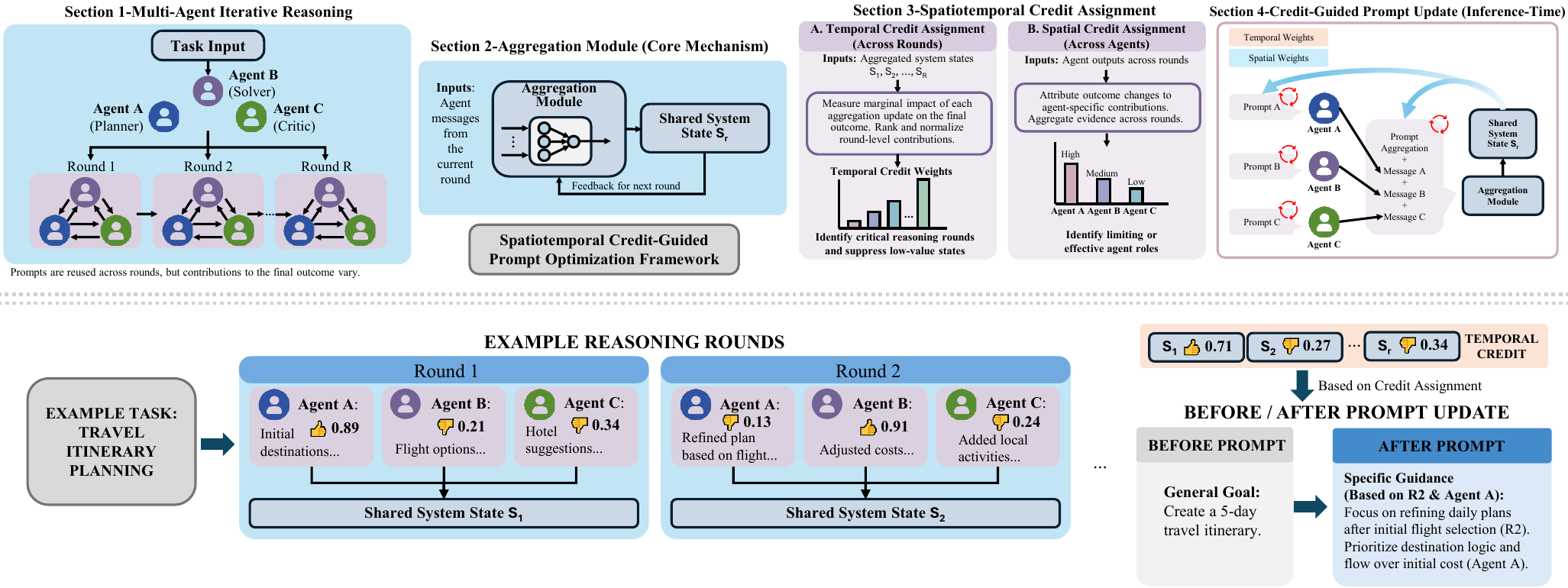}}
    \caption{Overview of the credit-guided prompt optimization pipeline. \textbf{Top:} a multi-agent, multi-round reasoning loop (planner/solver/critic) produces per-round messages that are aggregated into a shared system state $S_r$; an aggregation module feeds back to the next round. From the completed trajectory, we compute \emph{temporal credit} across rounds (identifying critical rounds) and \emph{structural/spatial credit} across agents (identifying effective or limiting roles). These credits then drive inference-time prompt updates, selectively refining the lowest-credit rounds/roles while keeping strong components fixed. \textbf{Bottom:} an example travel-itinerary task illustrates per-round agent outputs, the evolving shared state ($S_1, S_2, \ldots$), temporal credit weights, and a before/after prompt update that specializes guidance to the weak round/role.}
    \label{fig:methodology}
    \end{center}
    \vspace{-30pt}
\end{figure*}

However, optimizing the interaction dynamics of MAS presents a formidable challenge.
From an optimization perspective, an LLM-based MAS operates as a discrete, non-differentiable computation graph.
Unlike neural network training where backpropagation precisely attributes error to specific weights, MAS optimization suffers from the \textbf{Credit Assignment Problem (CAP)} in its most severe form: the supervisory signal is typically sparse (a single scalar score at the terminal state) and global (applying to the entire trajectory) \citep{SuttonBarto2018,Foerster2018COMA}.
Existing approaches often treat the entire system as a black box \citep{deng2022rlprompt, fernando2023promptbreeder, guo2023connecting}, applying derivative-free optimization or heuristic aggregation \citep{10.5555/3709347.3743784, ai2025majorityvotingllmaggregation}.

While effective for simple chains, these methods struggle in complex multi-round collaborations because they ignore the internal causal structure of the reasoning process.
Optimizing a system with $N$ agents over $R$ rounds essentially involves searching in a space of $\mathcal{O}(N \times R)$ coupled prompts. Without structural priors, this faces catastrophic variance and inefficient exploration \citep{zhou2024batch}.


In this work, we argue that making MAS optimization tractable requires imposing \textbf{structural constraints} on the computation graph to disentangle the error signal.
We propose two principled relaxations that transform the chaotic interaction graph into a structured optimization landscape.

First, state-space bottleneck via aggregation.
In a fully connected MAS, information flows diffusely, making it impossible to pinpoint \emph{when} reasoning collapsed.
We introduce an explicit \emph{Aggregation Module} at each round to summarize agent outputs into a unified state $S_t$.
This effectively models the interaction as a Markov Decision Process (MDP), where the aggregator creates a ``state bottleneck'' \citep{puterman1994mdp,mnih2015dqn}.
This architectural choice is not merely functional; it is a prerequisite for \textbf{Temporal Credit Assignment}, allowing us to evaluate the quality of consensus at discrete time steps independent of individual agent noise.

Second, stationary policy via parameter sharing \citep{GuptaEgorovKochenderfer2017,YuVeluVinitsky2022PPO}.
Optimizing unique prompts for every agent at every round creates an explosion of parameters.
We impose a \emph{Stationary Policy Constraint}, where each role (e.g., the Debater) shares the same system prompt across all interaction rounds.
This reduces the search space from trajectory-specific instructions to robust role definitions.
Crucially, this constraint enables reliable \textbf{Structural Credit Assignment}: by observing an agent's performance across multiple rounds, we can distinguish systematic role incompetence from transient stochastic errors.

Building on these structural insights, we propose a unified framework for \textbf{Temporal and Structural Credit Assignment}, as shown in Figure~\ref{fig:methodology}.
We formalize the MAS optimization problem as a bi-level objective depending on two orthogonal variable blocks: (i) \emph{Role Prompts} (defining agent behaviors) and (ii) \emph{Aggregation Prompts} (defining state transition protocols).
Instead of optimizing these jointly—which is prone to instability—we derive a novel Block Coordinate Descent (BCD) algorithm adapted for discrete prompt optimization \citep{Tseng2001BCD,Cai2023CyclicBCD}.
Our method alternates between refining role-specific instructions (fixing the aggregation logic) and refining the aggregation protocol (fixing the role behaviors).
To drive this descent, we utilize LLM-based critics to compute ``proxy gradients''—decomposed credit scores that guide targeted textual updates only to the bottleneck components.

Our contributions are summarized as follows:
1) We reformulate MAS prompt optimization by introducing state-space bottlenecks and stationary policy constraints. These inductive biases bridge the gap between sparse terminal rewards and dense, actionable feedback.
2) We propose a theoretically grounded optimization algorithm that alternates between structural (role) and temporal (aggregator) updates. This mimics Block Coordinate Descent in the discrete prompt space, ensuring stability and convergence compared to joint optimization baselines.
3) Numerical experiments demonstrate that our approach achieves higher task accuracy while reduces query complexity. By focusing updates on specific "weak link" roles or "phase transition" rounds, we avoid the inefficiency of indiscriminate global updates.

\section{Related Works}

\textbf{Prompt Optimization.} Several methods have been developed for optimizing prompts, ranging from learning parameter-efficient soft prompts to automatically searching for optimal discrete prompts using reinforcement learning, meta-optimization, and evolutionary algorithms.
Prompt-Tuning \citep{lester2021power} freezes the parameters of a pre-trained language model and learns task-specific soft prompts via end-to-end backpropagation.
RLPrompt \citep{deng2022rlprompt} optimizes discrete textual prompts through reinforcement learning with a policy network.
OPRO \citep{yang2023large} frames prompt optimization as an iterative black-box optimization problem.
EvoPrompt \citep{guo2023connecting} integrates evolutionary algorithms with large language models by employing the LLM to implement linguistically coherent crossover and mutation operations.
PromptBreeder \citep{fernando2023promptbreeder} implements a self-referential self-improvement mechanism for prompts using a genetic algorithm that co-evolves task-prompts and mutation-prompts.

\textbf{LLM-based MAS.} Current research on LLM-based multi-agent systems spans four key dimensions: system representation, optimization mechanisms, dynamic adaptation, and reasoning efficiency.
GPTSwarm \citep{zhang2024gptswarm} exemplifies offline structural learning via static post-training optimization with strong theoretical grounding in policy-gradient methods.
DyLAN \citep{liu2024dylan} represents unsupervised online selection, offering principled agent evaluation without labeled data and drawing theoretical connections to Shapley-value attribution.
EvoMAC \citep{li2024evomac} reflects environment-driven online adaptation, enabling test-time improvement from objective feedback and linking classical control-theoretic feedback to LLM generation.
MAS-GPT \citep{chen2024masgpt} illustrates generative offline learning, trading development-time cost for inference-time efficiency and cross-domain transfer.

\textbf{Credit Assignment.} The credit assignment problem refers to determining the contribution of each agent to the collectively obtained reward in a multi-agent system.
In reinforcement learning context, when an agent receives a reward, it is challenging to determine which actions should be credited or blamed for the outcome.
In multi-agent systems context, under the centralized training with decentralized execution paradigm, it is required to allocate contributions from joint decisions to individual agents. Value decomposition methods address this challenge by employing mixing networks to decompose the joint state-action value function into individual local observation-action value functions.
MATTRL \citep{hu2026collaborativemultiagenttesttimereinforcement} implements inference-time credit assignment via Difference Rewards for precise contribution attribution in multi-agent reasoning, while MAPRO \citep{zhang2025maprorecastingmultiagentprompt} employs training-time credit assignment for joint optimization of discussion and answer generation.

\section{Problem Formulation}\label{sec:problem-formulation}

We now formalize the multi-agent, multi-round prompting setting that our temporal and structural credit assignment targets. 
We specify the interaction protocol, notation, and the optimization variables, which will let us define per-round and per-role credit signals precisely. 
This section sets up the trajectory and objective used by the attribution method and the credit-guided prompt optimization procedure.

\paragraph{Multi-Agent, Multi-Round Prompting.}
We consider a multi-agent, multi-round LLM reasoning system with a fixed set of agents (roles)
$\mathcal{A}=\{a_1,\dots,a_N\}$ and $R$ interaction rounds.
The base LLM parameters are fixed; the optimization variables are the prompts used by each agent at each round.
Let $\phi_{i,t}$ denote the prompt fed to agent $a_i$ at round $t$.
We collect all prompts as
$\Phi \triangleq \{\phi_{i,t}\}_{i=1,t=1}^{N,R}$.

Given an input instance $x\sim\mathcal{D}$, the system runs for $R$ rounds.
At round $t$, each agent $a_i$ produces an utterance $u_{i,t}$ conditioned on $x$ and its prompt $\phi_{i,t}$:
\begin{equation}
u_{i,t} \sim p\!\left(\cdot \mid x,\phi_{i,t}\right), i\in\{1,\dots,N\},\ t\in\{1,\dots,R\}.
\label{eq:agent_gen}
\end{equation}
Let $\mathcal{U}_t \triangleq \{u_{1,t},\dots,u_{N,t}\}$ be the set of role utterances at round $t$.
An aggregation module summarizes $\mathcal{U}_t$ into a shared state $S_t = f_t(\mathcal{U}_t), \quad t\in\{1,\dots,R\}$, with an initial state $S_0$ fixed.

\paragraph{Prompt-as-Input and Output-as-Context.}
We define each round prompt as a composition of (i) a system instruction, (ii) the previous-round shared state,
and (iii) a role-specific template.
Concretely, for each agent $a_i$ and round $t$,
\begin{equation}
\phi_{i,t} = g_{i,t}\!\left(x, S_{t-1}; \theta_{i,t}\right),
\label{eq:prompt_compose}
\end{equation}
where $g_{i,t}(\cdot)$ is a (deterministic) prompt constructor and $\theta_{i,t}$ denotes its free text content
(the part we optimize).
This explicitly captures that each agent's \emph{output} at round $t-1$ becomes part of the
\emph{input} for round $t$.

\paragraph{Final-Round Scoring.}
After the final round, a terminal decision module produces the system output from $(x,S_R)$:
\begin{equation}
\hat{y} = J(x,S_R).
\label{eq:terminal_output}
\end{equation}
A task-specific scorer assigns a scalar score to the final output $s(x,\hat{y}) \in \mathbb{R}$.
Equivalently, the trajectory-level score can be written as
\begin{equation}
J(\tau;x) \triangleq s\!\left(x, j(x,S_R)\right),
\tau \triangleq \big(\mathcal{U}_1,S_1,\dots,\mathcal{U}_R,S_R\big).
\label{eq:traj_score}
\end{equation}

\paragraph{Optimization Objective.}
The basic prompt optimization problem is to maximize the expected final-round score:
\begin{equation}
\max_{\Phi}\;
\mathbb{E}_{x\sim\mathcal{D}}\,
\mathbb{E}_{\tau \sim p(\cdot \mid x;\Phi)}
\Big[ J(\tau;x) \Big].
\label{eq:objective}
\end{equation}
Since $J(\tau;x)$ is only observed at the end of the $R$-round interaction,
it depends on all upstream prompts $\{\phi_{i,t}\}$ through the coupled generation process
Eq.\eqref{eq:agent_gen}--Eq.\eqref{eq:terminal_output}.

\paragraph{Textual-Gradient Prompt Update.}
Because prompts are discrete text and the base LLM parameters are fixed,
we do not backpropagate numerical gradients through Eq.\eqref{eq:agent_gen}.
Instead, we convert the terminal score into a \emph{textual gradient} (natural-language feedback) \citep{Xiao2024VerbalizedML,Yuksekgonul2024TextGrad}
that specifies how to revise prompts:
\begin{equation}
\Delta_{i,t} = \mathcal{G}\!\left(x,\tau, J(\tau;x), i,t\right),
\label{eq:textual_grad}
\end{equation}
where $\mathcal{G}(\cdot)$ outputs a feedback string for prompt $(i,t)$.
Prompts are updated by applying the feedback to the current prompt text:
\begin{equation}
\phi_{i,t}^{(k+1)} = \mathcal{U}\!\left(\phi_{i,t}^{(k)}, \Delta_{i,t}^{(k)}\right),
\quad i=1,\dots,N,\ t=1,\dots,R,
\label{eq:prompt_update}
\end{equation}
where $k$ indexes optimization iterations and $\mathcal{U}(\cdot)$ is a text-edit operator (e.g., rewrite / insert constraints).
This completes the most basic definition of the optimization problem in a multi-agent, multi-round system,
where the final-round score is propagated back to prompts via textual gradients.

\section{Methodology}

Building on the naive objective in Eq.(\ref{eq:objective}), we make two relaxations to obtain a tractable optimization scheme.
First, we introduce a \emph{state-space bottleneck} by inserting a \emph{round-indexed} aggregation module (one aggregator per round), so round-level decisions become explicit and scorable.
Second, we impose \emph{parameter sharing} by instantiating a fixed set of role-specialized agents whose prompts are shared across rounds, so credits can be accumulated by role rather than by individual turns.
During optimization, we evaluate intermediate role outputs and per-round aggregation outputs with LLM-based critics to obtain \emph{proxy} signals for credit estimation. 
We use the terminal scorer $s(x,\hat{y})$ as the optimization objective, while LLM critics provide auxiliary signals to estimate role- and round-level credits for targeted prompt updates.
These critic signals are transformed into two forms of credit: (i) \emph{structural credit} over roles, used to identify and optimize
weak agents; and (ii) \emph{temporal credit} over rounds, implemented as a per-round aggregator credit score used to trigger targeted aggregation-prompt optimization.

\subsection{State-Space Bottleneck}
\label{sec:why-aggregation}

In MAS, the terminal score provides a single global optimization signal.
To update prompts, this signal must be expressed as a textual gradient.
Whether such feedback can be made actionable depends on the availability of a state-space bottleneck:
an explicit shared state that concentrates, rather than disperses, the information flow.

\paragraph{Without aggregation.}
If no aggregation is used, the final output is derived directly from the collection of all utterances
$\{u_{i,t}\}_{i,t}$.
In this case, the textual gradient can only be expressed as an unstructured global signal $\Delta = \mathcal{G}\!\left(x,\{u_{i,t}\}_{i,t}, J(\tau;x)\right)$.
Because there is no explicit bottleneck to anchor intermediate credit, the feedback cannot be aligned
with specific agents or interaction rounds, providing only coarse guidance for prompt updates.

\paragraph{With aggregation.}
By introducing an aggregation module, the system maintains explicit shared states $\{S_t\}$,
which instantiate a state-space bottleneck across rounds.
This enables textual gradients to be grounded on intermediate representations $\Delta_t = \mathcal{G}_t\!\left(x,S_t,S_R,J(\tau;x)\right)$, and further decomposed to individual agents:
\begin{equation}
\Delta_{i,t} = \mathcal{G}_{i,t}\!\left(x,u_{i,t},S_t,S_R,J(\tau;x)\right).
\end{equation}
In this way, aggregation provides the structural interface that makes textual gradients representable,
alignable, and decomposable across both temporal rounds and agent roles.

\paragraph{New optimization variables.}
Crucially, the aggregation module is itself prompt-driven, so introducing shared states also introduces
new decision variables: the aggregator prompts that control how $\mathcal{U}_t$ is summarized.
Let $\psi_t$ denote the aggregation prompt at round $t$, and write the state update as
$S_t = f_t(\mathcal{U}_t;\psi_t)$.
The optimization target therefore changes from role-prompt-only tuning to a joint objective over
role prompts and aggregation prompts:
\begin{equation}
\max_{\Phi,\Psi}\;
\mathbb{E}_{x\sim\mathcal{D}}\,
\mathbb{E}_{\tau \sim p(\cdot \mid x;\Phi,\Psi)}
\Big[ J(\tau;x) \Big],
\Psi \triangleq \{\psi_t\}_{t=1}^R .
\label{eq:objective_with_aggregation}
\end{equation}
This shift is important: optimizing aggregation can change the effective trajectory distribution
and thus the credit signals used to update role prompts.

\subsection{Parameter Sharing}
\label{sec:shared-prompt}

\begin{figure}[htb!]
    \centering
    \resizebox{\linewidth}{!}{%
    \begin{tikzpicture}[
        node distance=0.9cm,
        box/.style={rounded corners=2pt, draw=black!50, fill=black!2, inner xsep=5pt, inner ysep=4pt, align=center}
    ]
        \node[box] (obj0) {Naive objective\\ Eq.~\eqref{eq:objective}};
        \node[box, right=1.2cm of obj0] (obj1) {Add aggregation prompts\\ Eq.~\eqref{eq:objective_with_aggregation}};
        \node[box, right=1.2cm of obj1] (obj2) {Share role prompts\\ Eq.~\eqref{eq:shared_objective}};

        \draw[->, thick] (obj0) -- (obj1);
        \draw[->, thick] (obj1) -- (obj2);
    \end{tikzpicture}%
    }
    \caption{Evolution of the optimization objective: from the naive terminal-score objective, to introducing round-wise aggregation prompts, to enforcing shared role prompts.}
    \label{fig:objective-evolution}
\end{figure}
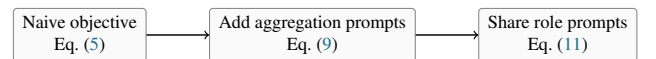

We also analyze a constrained regime with \emph{parameter sharing}, where each agent reuses a single
prompt across all interaction rounds. Formally, for each agent $a_i$, we impose
\begin{equation}
\phi_{i,1} = \phi_{i,2} = \cdots = \phi_{i,R} \triangleq \phi_i.
\label{eq:shared_prompt}
\end{equation}
This reduces the number of optimization variables, but increases the influence of each prompt
because it is applied repeatedly under different aggregated contexts.
Under this constraint, the optimization objective becomes
\begin{equation}
\max_{\{\phi_i\}_{i=1}^N,\Psi}
\;
\mathbb{E}_{x\sim\mathcal{D}}
\mathbb{E}_{\tau\sim p(\cdot\mid x;\{\phi_i\},\Psi)}
\big[ J(\tau;x) \big].
\label{eq:shared_objective}
\end{equation}
This is also our final optimization target.
See Fig.~\ref{fig:objective-evolution} for the overall evolution.
Although simplified, the problem remains non-trivial: each shared prompt $\phi_i$
affects the final score through multiple uses across evolving states $\{S_{t-1}\}$.
Effective optimization therefore requires aggregating feedback across rounds,
which naturally motivates temporal credit assignment under parameter sharing.

\subsection{Verbalized BCD}
\label{sec:critic-signals}

Having specified a new optimization objective, the next step is to design an algorithm that can effectively solve it.
Our objective provides supervision only at the end of the $R$-round interaction (Eq.~\eqref{eq:shared_objective}), leaving the intermediate rounds and roles without direct learning signals.
Crucially, terminal success does not imply that the final aggregation round (or speaker) is solely responsible; earlier rounds can be decisive or harmful.
We therefore introduce LLM-based critics as \emph{auxiliary} evaluators of intermediate utterances and aggregations, and translate their scores into temporal and structural credits.
These credits then drive a targeted \emph{block coordinate descent} procedure that alternates between updating low-credit role prompts and low-credit aggregation prompts, while keeping high-credit components fixed.

\paragraph{Credit Computing.}
For the \textit{structural dimension}, let $q_{i,t}\in[0,1]$ denote a normalized critic score for role $i$'s utterance $u_{i,t}$ at round $t$.
To account for interaction effects, we introduce a peer-view proxy $p_{i,t}\in[0,1]$:
we do \emph{not} require each agent to explicitly score others; instead, a single LLM-based judge conditions on the full set $\mathcal{U}_t$ and assesses each role's contribution to group reasoning.
We fuse these two signals as $c^{\text{role}}_{i,t} = \lambda\, q_{i,t} + (1-\lambda)\, p_{i,t}$ where $\lambda\in[0,1]$,
and aggregate across rounds to obtain a role-level structural credit $C^{\text{role}}_{i} = \frac{1}{R}\sum_{t=1}^R c^{\text{role}}_{i,t}$.
For the \textit{temporal dimension}, let $q^{\text{agg}}_t\in[0,1]$ be the critic score for the round-$t$ aggregation output $S_t$.
We define the temporal credit as $C^{\text{time}}_{t} = q^{\text{agg}}_t$.

\paragraph{Verbalized BCD over Prompt Blocks.}
Our optimization is a special instance of BCD over two prompt blocks:
(i) the \emph{structural} prompts that instantiate role behaviors, denoted by $\Phi^{\text{role}}$ (e.g., $\{\phi_i\}_{i=1}^N$ under the shared-prompt constraint in Eq.~\eqref{eq:shared_prompt}); and
(ii) the \emph{temporal} prompts that instantiate the round-indexed aggregation modules, denoted by $\Psi\triangleq\{\psi_t\}_{t=1}^R$.
Starting from $(\Phi^{\text{role},(0)},\Psi^{(0)})$, each outer iteration $k$ performs two alternating phases:

\begin{phasebox}{phaseA}
\paragraph{Phase A: optimize roles while fixing aggregation prompts.}
We keep $\Psi^{(k)}$ fixed, roll out trajectories, compute $\{C^{\text{role}}_i\}$, select a subset of low-credit roles (e.g., bottom-$K$ or below a threshold), and update only their prompts using textual gradients (Eq.~\eqref{eq:prompt_update}):
\begin{equation}
    \Phi^{\text{role},(k+1)} \leftarrow \textsc{RoleOpt}\big(\Phi^{\text{role},(k)};\, \Psi^{(k)},\, \{C^{\text{role}}_i\}\big).
\end{equation}
\end{phasebox}
\begin{phasebox}{phaseB}
\paragraph{Phase B: optimize aggregation prompts while fixing roles.}
We then keep $\Phi^{\text{role},(k+1)}$ fixed, roll out trajectories again (or reuse logged trajectories when applicable), compute $\{C^{\text{time}}_t\}$, select low-credit rounds, and update only their aggregation prompts:
\begin{equation}
    \Psi^{(k+1)} \leftarrow \textsc{AggOpt}\big(\Psi^{(k)};\, \Phi^{\text{role},(k+1)},\, \{C^{\text{time}}_t\}\big).
\end{equation}
\end{phasebox}

We repeat Phases A--B until a fixed budget is exhausted or the held-out performance saturates.
This alternating design makes the procedure implementable and stable: at each step we optimize one block while treating the other as a fixed environment, reducing unnecessary drift in the coupled multi-round interaction.

\begin{algorithm}[t]
\caption{BCD for Temporal and Structural Credit-Guided Prompt Optimization}
\label{alg:credit_bcd}

\KwIn{$\mathcal{D}$; fixed base LLM; rounds $R$; roles $\mathcal{A}=\{a_i\}_{i=1}^N$}
\KwOut{Optimized prompts $\{\phi_i\}_{i=1}^N$ and $\{\psi_t\}_{t=1}^R$}

\BlankLine
\textbf{Initialize:} $\{\phi_i^{(0)}\}$, $\{\psi_t^{(0)}\}$; $k\leftarrow 0$\;

\While{not converged}{

    \BlankLine
    \begin{algoblock}{phaseA}
    \tcp{\textbf{Step 1: Structural block update} (fix $\{\psi_t\}$)}
    Compute $\{C^{\text{role}}_i\}_{i=1}^N$\;
    Select low-credit roles $\mathcal{I}^{(k)}\subseteq\{1,\dots,N\}$\;
    \ForEach{$i\in\mathcal{I}^{(k)}$}{
        $\phi_i^{(k+1)} \leftarrow \mathcal{U}\!\big(\phi_i^{(k)},\Delta^{\text{role}}_i\big)$\;
    }
    Set $\phi_i^{(k+1)}\leftarrow \phi_i^{(k)}$ for $i\notin\mathcal{I}^{(k)}$\;
    \end{algoblock}

    \BlankLine
    \begin{algoblock}{phaseB}
    \tcp{\textbf{Step 2: Temporal block update} (fix $\{\phi_i\}$)}
    Compute $\{C^{\text{time}}_t\}_{t=1}^R$\;
    Select low-credit rounds $\mathcal{T}^{(k)}\subseteq\{1,\dots,R\}$\;
    \ForEach{$t\in\mathcal{T}^{(k)}$}{
        $\psi_t^{(k+1)} \leftarrow \mathcal{U}\!\big(\psi_t^{(k)},\Delta^{\text{time}}_t\big)$\;
    }
    Set $\psi_t^{(k+1)}\leftarrow \psi_t^{(k)}$ for $t\notin\mathcal{T}^{(k)}$\;
    \end{algoblock}

    \BlankLine
    $k\leftarrow k+1$\;
}
\Return $\{\phi_i^{(k)}\}_{i=1}^N,\{\psi_t^{(k)}\}_{t=1}^R$\;
\end{algorithm}

\section{Experiments}

To validate the proposed temporal and structural credit assignment and the resulting prompt optimization routine, we now turn to empirical evaluation. The following section specifies datasets, protocols, and baselines, and then quantifies how the credit signals translate into measurable gains and interpretability across benchmarks.

\subsection{Settings}
\label{sec:datasets}

We evaluate our proposed credit-guided prompt optimization on multiple-choice reasoning benchmarks, where each question has four options and the correct answer is selected from \{A, B, C, D\}. 
Models generate free-form text, and we deterministically extract the predicted option using a fixed parsing rule: the first occurrence of a standalone A/B/C/D token in the output. 
If no valid option is found, the prediction is considered invalid and counted as incorrect. 
These parsing rules are fixed prior to evaluation and remain constant across all runs. 
We report results on AQuA \citep{ling-etal-2017-program}, MedMCQA \citep{pal2022medmcqa}, GPQA \citep{rein2024gpqa}, and MMLU \citep{hendryckstest2021,hendrycks2021ethics}.

For each dataset, we randomly select a fixed \emph{optimization set} of 100 examples used exclusively for prompt search and optimization (i.e., updating role prompts and round-wise aggregator prompts). 
All reported metrics are evaluated on a larger, disjoint \emph{leave-out test set}, which is never accessed during optimization. 
The split is fixed across runs. 
No test examples are used for prompt search, gradient updates, credit assignment, or hyperparameter tuning. 
Accordingly, we emphasize \emph{no test-set exposure}, rather than zero-shot performance, on the reported test set.

Random seeds are fixed for sampling the optimization set and for any stochastic components in decoding or optimization. 
Unless otherwise stated, results are averaged over multiple runs with different seeds, and we report the mean accuracy along with standard deviation.
Prompt optimization is performed independently for each dataset. 
This constitutes dataset-specific prompt search over a fixed base model, not model fine-tuning. 
Cross-dataset generalization is not claimed unless explicitly evaluated.

\textbf{Models and Runtime.}
All debate agents are instantiated with open-weight instruction-following LLMs, allowing inspection and modification of prompts. 
We use Qwen2.5-7B-Instruct, LLaMA3-8B-Instruct, and Gemma-7B-Instruct for the main experiments, and decoding settings are kept identical across baselines and our method.

\textbf{Frameworks and Baselines.}
We evaluate our approach on several multi-agent frameworks, with a focus on \textbf{LLM-Debate} and \textbf{DyLAN}. 
In all settings, the interaction protocol (number of agents, number of rounds, and aggregator placement) is fixed. 
We compare against: 
(i) the unmodified baseline prompts, 
(ii) a black-box prompt optimization baseline using DSPy MIPRO (multi-stage instruction and demonstration optimization) \citep{opsahl-ong2024mipro} that performs agent-wise non-informed prompt search, and 
(iii) our credit-aware optimization.

\textbf{Training and Optimization Protocol.}
We follow a two-stage \emph{prompt-only} procedure; no model parameters are updated. 
First, DSPy (MIPROv2) is used solely for automated prompt initialization \citep{opsahl-ong2024mipro}. 
Second, fixed-budget prompt edits are performed using only the optimization split; the test split is never accessed.
\textit{(i) Structural optimization.} 
We evaluate each agent's final-round answer with \texttt{AgentTurnEval}, aggregate scores into a risk statistic, and deterministically select the two worst-performing roles. 
Only these role prompts are optimized using \texttt{RolePromptOpt} with summarized error diagnosis.
\textit{(ii) Temporal optimization.} 
Temporal credit is assigned to per-round aggregators using the EMA update described in Section~3. 
Credit is updated only on informative failures (aggregator wrong but at least one agent correct). 
When the credit of round $t$ drops below a threshold and sufficient failures are observed, only that round's aggregator prompt is optimized.
All hyperparameters are fixed across datasets.



\subsection{Main Results and Analysis}

Based on the analyses in this section, the experiments aim to answer four research questions:
\textbf{RQ1:} Does credit-guided prompt optimization consistently improve accuracy across datasets and models compared with unoptimized prompts and a black-box baseline?
\textbf{RQ2:} How much do structural and temporal optimization each contribute, and does their combination yield the strongest and most consistent gains?
\textbf{RQ3:} Is the optimization process efficient and stable (convergence speed and variance), and how do key hyperparameters affect performance and stability?
\textbf{RQ4:} Do credit signals align with error types and outcome shifts, providing interpretable diagnostics and actionable guidance for follow-up optimization?

\subsection{RQ1: Effectiveness}

Table~\ref{tab:main_results} summarizes the performance of our credit-guided prompt optimization compared to two baselines: the unmodified prompts and a black-box optimization baseline. 
Across all datasets and models, our method consistently improves accuracy. 
For instance, on MedMCQA with LLaMA3-8B, our approach improves accuracy by 7.0\% relative to the baseline, while on GPQA with Qwen2.5-7B the gain is 2.1\%. 
Even on datasets with higher initial accuracy, such as AQuA and MMLU, we observe consistent improvements. 
Standard deviations across multiple runs remain low, indicating stable and reproducible performance. 
These results confirm that integrating structural and temporal credit signals effectively guides prompt optimization in multi-agent settings.
\begin{phasebox}{phaseC}
\textbf{RQ1 Answer.} Credit-guided prompt optimization delivers broad, reliable accuracy gains across datasets and model families, outperforming both the unoptimized prompts and the black-box baseline, which supports its generalizability and robustness.
\end{phasebox}

\begin{table*}[htb!]
\caption{Main results on multiple-choice reasoning benchmarks. The main number is the mean accuracy, and the right-side annotation reports the delta relative to the corresponding MAS baseline: ``-'' marks the baseline, $\uparrow$ denotes an increase, and $\downarrow$ denotes a decrease. Cell colors follow a red--white--green gradient from negative to zero to positive deltas, and the lower number in each cell is the standard deviation.}
\label{tab:main_results}
\def\cboxwgt{9mm}
\def\sp{3mm}
\begingroup\small
\begin{tabular*}{\linewidth}{@{\extracolsep{\fill}} c c c c c @{\hspace{1.5em}} c c c}
    \toprule
     & & \multicolumn{3}{c}{DyLAN} & \multicolumn{3}{c}{Debate}  \\
    \cmidrule(lr){3-5} \cmidrule(lr){6-8}
    Domain & Model & baseline & optimized & ours & baseline & optimized & ours \\
    \midrule
    \multirow{3}{*}[-1mm]{\parbox[m][][c]{1.5cm}{\centering MedMCQA}} 
& LLaMA3-8B  
& \cbox{54.13}{2.3936}{0}{-} 
& \cbox{61.75}{3.3040}{7.63}{\uparrow 7.63} 
& \cboxours{61.13}{1.0308}{7.00}{\uparrow 7.00}
& \cbox{55.13}{3.2243}{0}{-} 
& \cbox{61.63}{2.1360}{6.50}{\uparrow 6.50}
& \cboxours{64.63}{1.7970}{9.50}{\uparrow 9.50}  \\ [\sp]
& Qwen2.5-7B
& \cbox{51.88}{2.3229}{0}{-} 
& \cbox{53.88}{1.6520}{2.00}{\uparrow 2.00} 
& \cboxours{55.75}{1.9365}{3.88}{\uparrow 3.88}
& \cbox{54.50}{1.0000}{0}{-} 
& \cbox{54.88}{3.3510}{0.38}{\uparrow 0.38}
& \cboxours{55.50}{4.3205}{1.00}{\uparrow 1.00}  \\ [\sp]
& Gemma-7B
& \cbox{19.13}{1.1087}{0}{-} 
& \cbox{19.38}{0.9465}{0.25}{\uparrow 0.25} 
& \cboxours{21.33}{2.4664}{2.21}{\uparrow 2.21}
& \cbox{26.75}{3.1820}{0}{-} 
& \cbox{27.50}{2.4833}{0.75}{\uparrow 0.75}
& \cboxours{28.50}{2.5981}{1.75}{\uparrow 1.75} \\
    \midrule
    \multirow{3}{*}[-1mm]{\parbox[m][][c]{1.5cm}{\centering GPQA}} 
& LLaMA3-8B  
& \cbox{27.42}{2.2728}{0}{-} 
& \cbox{29.61}{3.0873}{2.19}{\uparrow 2.19} 
& \cboxours{28.11}{1.8779}{0.69}{\uparrow 0.69}
& \cbox{39.75}{1.7816}{0}{-} 
& \cbox{35.02}{1.5535}{-4.73}{\downarrow 4.73}
& \cboxours{36.75}{4.4439}{-2.30}{\downarrow 2.30} \\ [\sp]
& Qwen2.5-7B
& \cbox{32.83}{2.3308}{0}{-} 
& \cbox{35.02}{2.0956}{2.19}{\uparrow 2.19} 
& \cboxours{32.37}{1.6581}{-0.46}{\downarrow 0.46}
& \cbox{32.26}{1.4546}{0}{-} 
& \cbox{33.07}{1.3212}{0.81}{\uparrow 0.81}
& \cboxours{35.83}{0.8752}{3.57}{\uparrow 3.57} \\ [\sp]
& Gemma-7B
& \cbox{11.52}{0.4600}{0}{-} 
& \cbox{12.29}{0.5312}{0.77}{\uparrow 0.77} 
& \cboxours{11.98}{0.6505}{0.46}{\uparrow 0.46}
& \cbox{12.56}{1.3778}{0}{-} 
& \cbox{17.75}{3.5850}{5.19}{\uparrow 5.19}
& \cboxours{18.67}{1.6334}{6.11}{\uparrow 6.11} \\
    \midrule
    \multirow{3}{*}[-1mm]{\parbox[m][][c]{1.5cm}{\centering AQuA}} 
& LLaMA3-8B  
& \cbox{51.15}{3.0376}{0}{-} 
& \cbox{55.46}{2.5500}{4.31}{\uparrow 4.31} 
& \cboxours{56.32}{1.6263}{5.17}{\uparrow 5.17}
& \cbox{53.16}{3.6784}{0}{-} 
& \cbox{52.01}{4.3354}{-1.15}{\downarrow 1.15}
& \cboxours{52.30}{9.4087}{-0.86}{\downarrow 0.86}  \\ [\sp]
& Qwen2.5-7B
& \cbox{83.05}{1.7250}{0}{-} 
& \cbox{85.35}{1.7250}{2.30}{\uparrow 2.30} 
& \cboxours{83.91}{1.3279}{0.86}{\uparrow 0.86}
& \cbox{83.05}{2.8750}{0}{-} 
& \cbox{81.04}{1.4846}{-2.01}{\downarrow 2.01}
& \cboxours{83.62}{3.3031}{0.58}{\uparrow 0.58}  \\ [\sp]
& Gemma-7B
& \cbox{18.72}{2.9480}{0}{-} 
& \cbox{22.42}{0.8132}{3.70}{\uparrow 3.70} 
& \cboxours{23.57}{0.6640}{4.85}{\uparrow 4.85}
& \cbox{18.62}{1.4994}{0}{-} 
& \cbox{22.61}{1.3279}{3.99}{\uparrow 3.99}
& \cboxours{27.59}{2.3000}{8.97}{\uparrow 8.97} \\
    \midrule
    \multirow{3}{*}[-1mm]{\parbox[m][][c]{1.5cm}{\centering MMLU}} 
& LLaMA3-8B  
& \cbox{56.43}{1.0820}{0}{-} 
& \cbox{56.20}{0.5952}{-0.23}{\downarrow 0.23} 
& \cboxours{56.30}{0.7407}{-0.13}{\downarrow 0.13}
& \cbox{68.78}{1.0444}{0}{-} 
& \cbox{69.33}{2.6207}{0.55}{\uparrow 0.55}
& \cboxours{74.77}{2.8646}{6.00}{\uparrow 6.00} \\ [\sp]
& Qwen2.5-7B
& \cbox{65.35}{1.1045}{0}{-} 
& \cbox{66.38}{1.1386}{1.03}{\uparrow 1.03} 
& \cboxours{67.48}{0.6135}{2.13}{\uparrow 2.13}
& \cbox{61.58}{1.3371}{0}{-} 
& \cbox{61.95}{1.9265}{0.38}{\uparrow 0.38}
& \cboxours{63.38}{2.0126}{1.80}{\uparrow 1.80}  \\ [\sp]
& Gemma-7B
& \cbox{29.40}{0.3162}{0}{-} 
& \cbox{30.40}{0.4733}{1.00}{\uparrow 1.00} 
& \cboxours{32.00}{0.2828}{2.60}{\uparrow 2.60}
& \cbox{28.25}{4.8946}{0}{-} 
& \cbox{33.73}{3.7754}{5.48}{\uparrow 5.48}
& \cboxours{33.60}{2.2627}{5.35}{\uparrow 5.35} \\
    \bottomrule
\end{tabular*}
\endgroup
\end{table*}

\subsection{RQ2: Ablations}

To understand the source of improvements, we perform ablation experiments. 
We isolate the contributions of structural and temporal components:
\textbf{Role-only optimization (structural):} optimizing only the worst-performing role prompts selected by the risk statistic.
\textbf{Aggregator-only optimization (temporal):} optimizing only the low-credit round aggregators while freezing role prompts.

\begin{figure}[ht]
    \centering
    \includegraphics[width=\columnwidth]{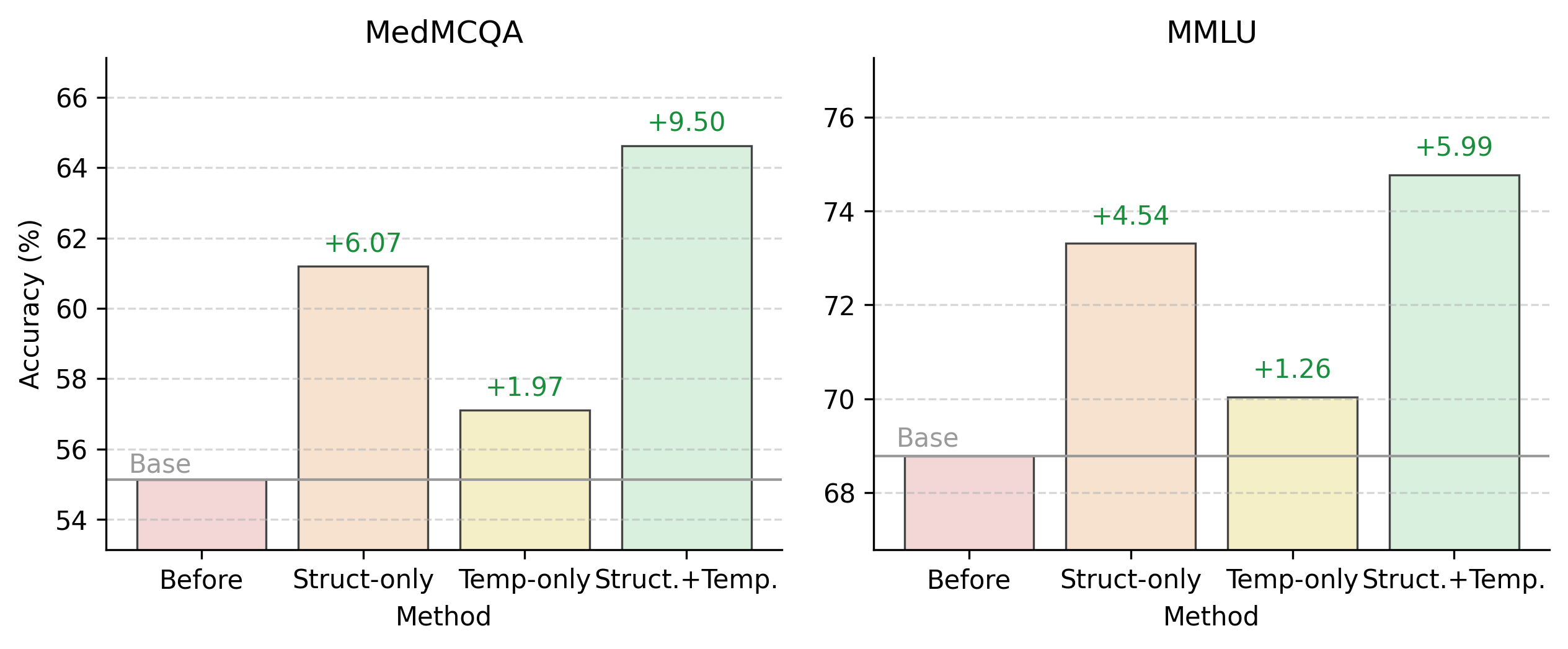}
    \caption{LLaMA3-8B Debate results on MedMCQA and MMLU: the combined temporal+structural credit-guided optimization yields the strongest gains, with structural-only and temporal-only providing smaller improvements over the baseline.}
    \label{fig:medmcqa_mmlu}
\end{figure}

Figure~\ref{fig:medmcqa_mmlu} visualizes the LLaMA3-8B Debate results on MedMCQA and MMLU, highlighting the consistent gains from credit-guided prompt optimization.
Both components independently improve accuracy relative to the baseline. 
Structural optimization contributes more on role-sensitive datasets (e.g., MedMCQA), whereas temporal optimization is most effective on datasets where aggregator consolidation is critical (e.g., GPQA). 
The combination consistently achieves the highest performance.  

We further analyze temporal optimization at the round level. Restricting updates to one round $t$ at a time reveals which rounds are most error-prone. 
Early rounds often dominate failures, indicating that losing key evidence early can limit overall accuracy, while late rounds mainly affect final consolidation. 
These findings provide actionable guidance on where optimization should focus.
\begin{phasebox}{phaseC}
\textbf{RQ2 Answer.} The ablations show complementary effects: role-only (structural) updates yield larger gains on role-sensitive datasets such as MedMCQA, while aggregator-only (temporal) updates are more beneficial on consolidation-heavy tasks such as GPQA. Combining both consistently gives the best accuracy across MedMCQA and MMLU, outperforming either component alone.
\end{phasebox}

\subsection{RQ3: Sensetivity}

\begin{figure}[ht]
    \centering
    \includegraphics[width=.8\columnwidth]{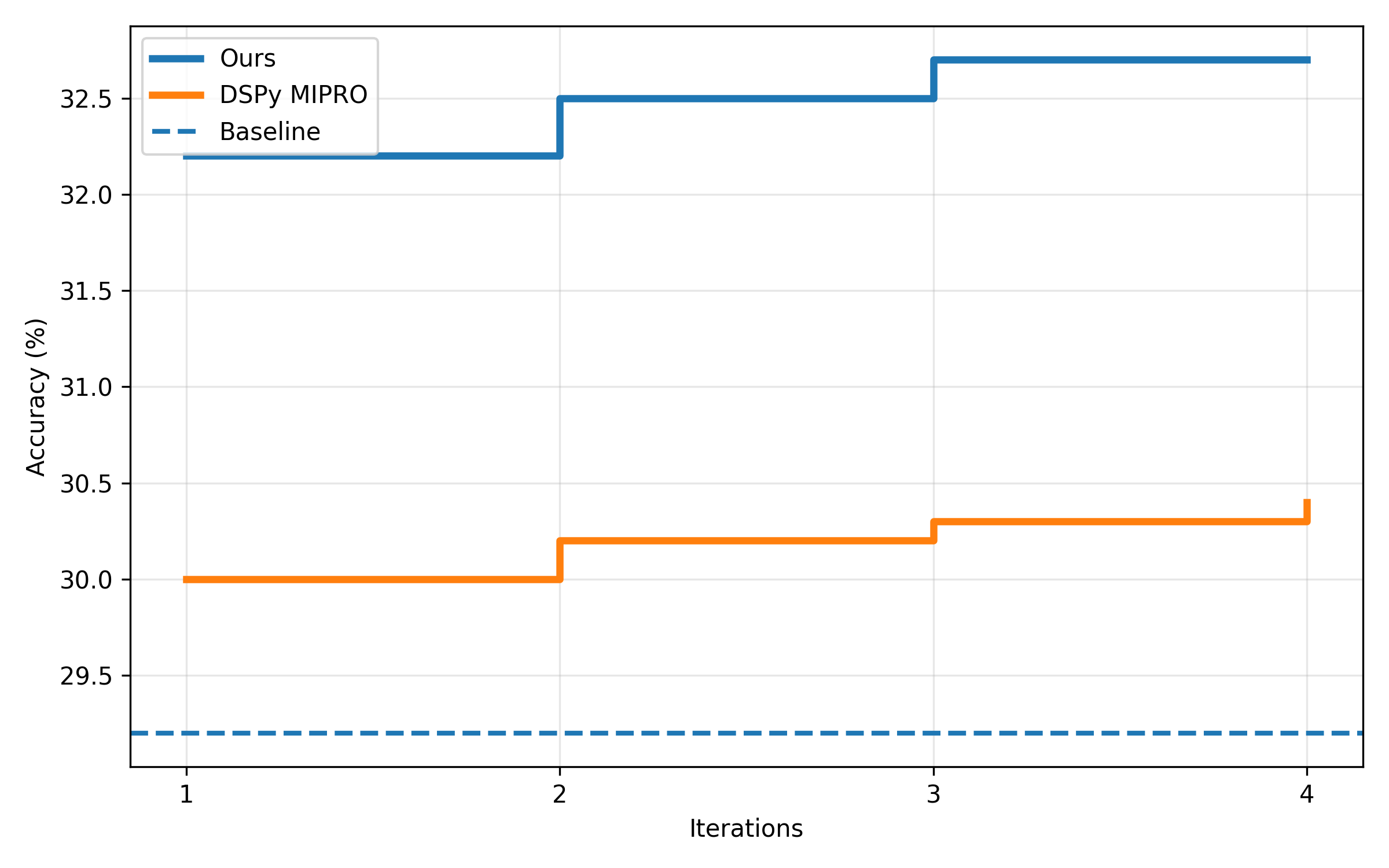}
    \caption{Accuracy convergence over optimization iterations; ours converges faster and to higher accuracy than DSPy MIPRO and the baseline.}
    \label{fig:it}
\end{figure}

We evaluate performance across optimization iterations to assess efficiency and stability. 
Figure~\ref{fig:it} shows the accuracy trajectories across iterations, highlighting faster gains for our method versus DSPy MIPRO and the baseline. 
Credit-guided optimization converges rapidly within a few iterations and maintains low variance across runs. 
In contrast, the black-box baseline converges more slowly and exhibits higher variance, reflecting inefficient exploration. 

We also investigate hyperparameter sensitivity by varying the temporal credit threshold $\tau$ and buffer size $m$. 
Lower $\tau$ or larger $m$ results in conservative updates, preserving stability, while higher $\tau$ or smaller $m$ allows more aggressive updates, increasing the risk of performance drift. 
Monitoring credit trajectories $\alpha_t$ and test accuracy variance confirms that our recommended settings achieve a favorable stability–plasticity balance.
\begin{phasebox}{phaseC}
\textbf{RQ3 Answer.} Building on the consistent gains in RQ1 and the complementary effects in RQ2, the optimization is efficient because it targets the most influential components, achieving improvements with only a few focused iterations. Stability follows from this targeted selection, while $\tau$ and $m$ provide explicit control over the trade-off between cautious refinement and faster but riskier updates.
\end{phasebox}

\subsection{RQ4: Interpretability}

\begin{table}[ht]
    \caption{Distribution of prediction shifts before and after optimization.
Compared with DSPu(MIPRO), our method yields more incorrect$\rightarrow$correct repairs and fewer correct$\rightarrow$incorrect regressions, indicating that credit signals act as structured diagnostics rather than global perturbations.}
    \label{tab:shift}
    \centering
    \renewcommand{\arraystretch}{1.15}
    \begin{tabular}{lcc}
        \toprule
        Shift & DSPy(MIPRO) (\%) & Ours (\%) \\
        \midrule
        X$\rightarrow$X & 38.00 & 34.50 \\
        \rowcolor{red!12}
        \checkmark$\rightarrow$X & 7.00 & 5.00 \\
        \rowcolor{green!12}
        X$\rightarrow$\checkmark & 7.00 & 11.00 \\
        \checkmark$\rightarrow$\checkmark & 48.00 & 49.50 \\
        \bottomrule
    \end{tabular}
\end{table}

To validate that optimization aligns with failure diagnosis, we analyze outcome shifts and error types. 
Compared with black-box edits, credit-guided optimization reduces regressions (correct$\to$incorrect) and increases repairs (incorrect$\to$correct). 
Per-category accuracy shows that role-specific optimizations mainly improve categories flagged as failing by the critic, and temporal credit $\alpha_t$ correlates with aggregator reliability rather than mere signal presence. 
Together, these analyses confirm that structural and temporal credit signals provide both mechanistic interpretability and practical guidance for follow-up experiments.
\begin{phasebox}{phaseC}
\textbf{RQ4 Answer.} Taken together with RQ1--RQ3, the credit signals function as actionable diagnostics: they pinpoint which rounds and roles actually drive outcome changes, explain when gains persist or regress under tuning, and indicate where the next optimization pass should focus.
\end{phasebox}

\section{Closing Remarks}

In this paper, we introduce a test-time temporal and structural credit assignment framework that decomposes multi-agent LLM trajectories along rounds and roles, enabling component-level attribution without altering inference dynamics or model parameters. 
Across the experiments, the resulting credit signals reveal pronounced contribution imbalance and identify the small set of stages and agents that consistently constrain outcomes. 
Leveraging these signals, credit-guided prompt optimization focuses updates on weak components and delivers reliable gains across benchmarks and model families, reducing wasteful global changes while improving effectiveness.

These findings position temporal and structural attribution as a practical diagnostic and optimization tool for multi-agent reasoning systems. 
At the same time, the framework remains limited to completed trajectories and relies on evaluative signals; temporal attribution can be computationally heavy for long horizons; and the current instantiation assumes fixed roles with centralized aggregation. Future directions include integrating credit signals into online or inference-time adaptation, extending attribution to weakly supervised settings via self-consistency or consensus critics, and applying the framework to more dynamic coordination regimes such as hierarchical, tool-augmented, or open-ended environments.

\clearpage
\newpage

\bibliography{main}

\newpage

\onecolumn

\appendix
\section{Appendix Overview}
This appendix provides supplemental details to support reproducibility and analysis.
We use standard appendix numbering (e.g., Table~A.1, Figure~A.1, Algorithm~A.1) and keep all references consistent with the main text.

\section{Evaluation and Prompt Optimisation Prompts}

The prompts used for agent evaluation, diagnosis and role prompt optimisation are presented below to aid reproducibility.  These prompts are treated as experimental protocols that can be referenced in the main text.

\subsection{Agent Turn Evaluation Prompt}

\begin{tcblisting}{
  title=Agent Turn Evaluation Prompt,
  colback=mygreen!5!white,
  colframe=mygreen!75!black,
  colbacktitle=mygreen!20!white,
  coltitle=black,
  fonttitle=\bfseries,
  listing only,
  breakable
}
AGENT_TURN_EVAL_PROMPT = """
You are an experienced question evaluation specialist and failure-analysis prompt engineer.

You are given:
1) A multiple-choice question with options A, B, C, and D.
2) The gold correct answer.
3) The final answer produced by ONE debating agent.

Your task is NOT to solve the question again.

Your task is to evaluate this agent’s final answer independently and assign a score that reflects the quality of its reasoning and decision-making.

You must:
- Determine whether the final answer is correct.
- If incorrect, identify the primary reason for failure using a predefined category.
- If correct, assess whether the reasoning is robust or fragile.
- Assign a numerical score that can be accumulated across many questions.

--------------------------------
Failure pattern taxonomy (choose EXACTLY ONE):

You MUST select exactly one of the following labels.
Do NOT invent new labels.

- DOMAIN_MISMATCH  
  The agent reasons from an inappropriate domain or role (e.g., ethical, economic, or technical framing instead of task-relevant reasoning).

- KNOWLEDGE_DEFICIT  
  The agent lacks or misuses core domain knowledge, leading to factual or conceptual errors.

- MISINTERPRET_QUESTION  
  The agent misunderstands key constraints, conditions, or intent of the question.

- INCOMPLETE_REASONING  
  The agent’s reasoning is partially correct but missing critical logical steps or justification.

- OVERGENERALIZATION  
  The agent applies generic patterns or heuristics without adequately considering case-specific details.

- MISALIGNED_OBJECTIVE  
  The agent answers a different question than what is being asked (e.g., treatment vs diagnosis).

- INSUFFICIENT_JUSTIFICATION  
  The conclusion may be correct, but the reasoning is weak, shallow, or insufficiently supported.

- RANDOM_OR_UNGROUNDED  
  The answer appears arbitrary, speculative, or not grounded in the provided information.

- NONE  
  The answer is correct and the reasoning is sound.

--------------------------------
Scoring rules:
- Scores must be integers from 0 to 5.
- 5 = Correct answer with strong, well-justified reasoning.
- 3–4 = Correct answer but with weak, incomplete, or risky reasoning.
- 1–2 = Incorrect answer due to reasoning or judgment errors.
- 0 = Incorrect answer due to fundamental misunderstanding or systematic reasoning failure.

Consistency rules:
- If the answer is correct, failure_pattern MUST be "NONE".
- If the answer is incorrect, failure_pattern MUST NOT be "NONE".

Output format:
- Final answer correctness: correct / incorrect
- Primary failure or risk pattern: <one of the predefined labels>
- Brief explanation: 1–2 sentences focused on reasoning quality
- Score: <integer 0–5>

Additional rules:
1. Do NOT restate the full question or options.
2. Do NOT compare with other agents.
3. Do NOT suggest prompt changes explicitly.
4. Focus on issues that could be mitigated by improving the agent’s prompt or debate behavior.
5. Be concise, consistent, and scoring-oriented.
"""
\end{tcblisting}

\subsection{Agent Diagnosis Prompt}

\begin{tcblisting}{
  title=Agent Diagnosis Prompt,
  colback=myred!4!white,
  colframe=myred!70!black,
  colbacktitle=myred!18!white,
  coltitle=black,
  fonttitle=\bfseries,
  listing only,
  breakable
}
AGENT_DIAGNOSIS_PROMPT = """
You are an attribution analyst for a multi-agent reasoning system.

Your role is to analyze summarized error information produced by a single agent
and generate a concise attribution summary of the agent’s systematic failure characteristics.

The input you receive is a structured summary grouped by failure types.
For each failure type, the input provides:
- how frequently this failure occurred, and
- a small number of representative example explanations.

You should treat the frequency information as an indicator of how systematic
and dominant each failure pattern is.
The example explanations are only illustrative signals and do NOT represent all errors.

You are NOT responsible for fixing the errors or rewriting the agent’s role prompt.
Your task is strictly to identify and summarize why the errors occurred.

When producing the attribution summary:

- Focus on dominant and recurring failure patterns, prioritizing those with higher frequency.
- Identify shared reasoning weaknesses, perspective mismatches, or systematic misalignments.
- Abstract away from individual questions, examples, or surface details.
- Do NOT repeat or quote raw explanations.
- Do NOT enumerate failure types or counts explicitly.
- Do NOT include task-specific facts or domain knowledge.

Your summary should capture, at an appropriate level of abstraction:
- role or perspective mismatches,
- reasoning or interpretation deficiencies,
- knowledge usage or grounding issues,
if they are reflected in the summarized failures.

The output should be concise, structured in natural language,
and suitable for downstream prompt refinement modules.

Output only the attribution summary.
Do not include analysis, bullet points, statistics, or recommendations.
"""
\end{tcblisting}

\subsection{Role Prompt Optimisation Prompt}

\begin{tcblisting}{
  title=Role Prompt Optimisation Prompt,
  colback=myblue!5!white,
  colframe=myblue!70!black,
  colbacktitle=myblue!18!white,
  coltitle=black,
  fonttitle=\bfseries,
  listing only,
  breakable
}
ROLE_PROMPT_OPTIMIZE = """
You are a prompt refinement module for a multi-agent reasoning system.

Your task is to correct and reconstruct an agent's role prompt based on
aggregated evaluation failures. The original role prompt may contain
invalid assumptions, missing constraints, or misleading reasoning guidance.
Do NOT assume the original prompt is correct.

Use the failure summary to infer systematic issues in three aspects:
(1) role and perspective alignment,
(2) reasoning and interpretation discipline,
(3) knowledge use and grounding.

Follow these principles implicitly:

1. Role and objective correction:
   - If failures indicate domain mismatch or objective misalignment,
     remove or correct the role perspective, focus, or priorities
     that cause the agent to reason from an inappropriate viewpoint
     or answer the wrong question.

2. Reasoning discipline reconstruction:
   - If failures indicate misinterpretation, incomplete reasoning,
     overgeneralization, or weak justification,
     introduce clearer reasoning requirements such as
     careful question interpretation, constraint checking,
     step-by-step reasoning, and explicit justification.

3. Knowledge use and grounding control:
   - If failures indicate knowledge deficits or ungrounded responses,
     strengthen guidance on using only relevant, task-appropriate knowledge
     and avoiding speculative or unsupported conclusions.

Constraints:
- Do not preserve incorrect assumptions from the original prompt.
- Do not add task-specific facts or external domain knowledge.
- Do not overfit to individual examples; address systematic behavior only.
- Keep the reconstructed prompt concise and suitable for debate-based interaction.

Output only the reconstructed role prompt.
Do not include analysis, explanations, or failure labels.
"""
\end{tcblisting}

\section{Extended Analysis of the DyLAN Baseline on MMLU (Example)}
\label{sec:appendix-log-analysis}

To aid reproducibility and provide deeper insight into system behavior, we analyzed the evaluation log file \texttt{8080\_dylan\_mmlu\_2026-02-19\_21-20-23.txt}. This file contains 500 entries, each recording the final answer of a multi-agent debate, the evaluator’s judgment of correctness, a failure-pattern label, and a quality score. The analysis below summarizes the distributions of these values.

\section{Extended Analysis of Optimized DyLAN on MMLU (Example)}
\label{sec:appendix-optimised-experiment}

In addition to the baseline analysis, we examined the log file from the optimized DyLAN system (\texttt{8080\_dylan\_mmlu\_2026-02-19\_21-20-44.txt}). This experiment applied an optimization method to the role prompts before evaluation. The log comprises 500 multiple-choice questions from the MMLU dataset.

\subsection{Overall Prediction Statistics}
Table~\ref{tab:exp-overall} summarizes the aggregate performance of the optimized system. Out of 500 questions, the system answered 279 correctly (55.8\% accuracy). The mean evaluation score was 3.38 on a 0--5 scale, comparable to the baseline run. Thus, despite optimization, there was no substantial improvement in overall accuracy or score.

\begin{table}[htb!]
    \caption{Overall statistics for the optimised DyLAN system on MMLU.}
    \label{tab:exp-overall}
    \centering
    \begin{tabular}{lll}
        \toprule
        Metric & Value & Description \\
        \midrule
        Total predictions & 500 & Number of evaluated questions \\
        Correct predictions & 279 (55.8\%) & Final answers matching the gold answer \\
        Incorrect predictions & 221 (44.2\%) & Final answers that were wrong \\
        Average evaluation score & 3.38 & Mean score on a 0–5 scale \\
        \bottomrule
    \end{tabular}
\end{table}

\subsection{Distribution Across Debate Rounds}
The optimization encouraged earlier convergence: 81.6\% of questions were answered after one round, while only 3.4\% of questions reached the third round (Table~\ref{tab:exp-round}). The average score dropped sharply from round~1 to round~2 (3.59 to 2.27) and increased again for the few questions that reached round~3 (3.12).

\begin{table}[htb!]
    \caption{Number of questions and average score by debate round for the optimised system.}
    \label{tab:exp-round}
    \centering
    \begin{tabular}{rrrr}
        \toprule
        Round & Count & Proportion & Average score \\
        \midrule
        1 & 408 & 81.6\% & 3.59 \\
        2 & 75 & 15.0\% & 2.27 \\
        3 & 17 & 3.4\% & 3.12 \\
        \bottomrule
    \end{tabular}
\end{table}

\subsection{Failure Pattern Distribution}
Table~\ref{tab:exp-failure} reports the frequency and average score of each failure pattern. As with the baseline run, \texttt{KNOWLEDGE\_DEFICIT} and \texttt{MISINTERPRET\_QUESTION} remain the most common error modes, collectively accounting for nearly 38\% of all predictions. Additional errors such as \texttt{INCOMPLETE\_REASONING} and \texttt{OVERGENERALIZATION} appear more often than in the baseline, and a single \texttt{RANDOM\_OR\_UNGROUNDED} error is observed.

\begin{table}[htb!]
    \caption{Failure patterns and average scores for the optimised system.  ``NONE'' denotes correct answers.}
    \label{tab:exp-failure}
    \centering
    \resizebox{.9\textwidth}{!}{%
    \begin{tabular}{lrrrl}
        \toprule
        Failure pattern & Count & Share & Avg. score & Interpretation \\
        \midrule
        NONE (correct) & 279 & 55.8\% & 5.00 & Correct answer with robust reasoning \\
        KNOWLEDGE\_DEFICIT & 123 & 24.6\% & 1.33 & Missing or misused domain knowledge \\
        MISINTERPRET\_QUESTION & 65 & 13.0\% & 1.32 & Misunderstanding the question’s intent or constraints \\
        INCOMPLETE\_REASONING & 11 & 2.2\% & 1.45 & Partial reasoning missing critical steps \\
        MISALIGNED\_OBJECTIVE & 8 & 1.6\% & 1.38 & Answering an off‑target question \\
        OVERGENERALIZATION & 6 & 1.2\% & 1.33 & Generic heuristics without context \\
        INSUFFICIENT\_JUSTIFICATION & 4 & 0.8\% & 1.50 & Plausible conclusion but weak justification \\
        DOMAIN\_MISMATCH & 3 & 0.6\% & 1.00 & Reasoning from an inappropriate domain \\
        RANDOM\_OR\_UNGROUNDED & 1 & 0.2\% & 1.00 & Arbitrary or ungrounded answer \\
        \bottomrule
    \end{tabular}
    }
\end{table}

\subsection{Round–Error Cross Analysis}
Table~\ref{tab:exp-cross} cross-tabulates failure patterns across debate rounds. Most correct responses occur in the first round. Knowledge deficits remain prevalent across rounds, whereas misinterpretation errors are more prominent in the third round. The distribution suggests that the optimization did not fully mitigate fundamental knowledge gaps.

\begin{table}[htb!]
    \caption{Number of predictions by round and failure pattern for the optimised system.}
    \label{tab:exp-cross}
    \centering
    \footnotesize
    \setlength{\tabcolsep}{3pt}
    \resizebox{\textwidth}{!}{%
    \begin{tabular}{r|rrrrrrrrr}
        \toprule
        & NONE & \makecell[c]{KNOWL.\newline DEFICIT} & \makecell[c]{MISINT.\newline QUESTION} & \makecell[c]{MISALIGN.\newline OBJ.} & \makecell[c]{INCOMP.\newline REASON.} & \makecell[c]{DOMAIN\newline MISM.} & \makecell[c]{OVERGEN.} & \makecell[c]{INSUFF.\newline JUST.} & \makecell[c]{RANDOM\newline UNGR.} \\
        \midrule
        Round~1 & 251 & 92 & 42 & 6 & 7 & 2 & 4 & 3 & 1 \\
        Round~2 & 20 & 29 & 17 & 2 & 3 & 1 & 2 & 1 & 0 \\
        Round~3 & 8 & 2 & 6 & 0 & 1 & 0 & 0 & 0 & 0 \\
        \bottomrule
    \end{tabular}%
    }
\end{table}
\end{document}